\documentclass[review]{elsarticle}
\pdfoutput=1
\usepackage{graphicx}
\usepackage{algorithm}
\usepackage{algorithmic}
\usepackage{float}
\usepackage[hidelinks]{hyperref}
\usepackage{amsthm}
\usepackage{verbatim}
\usepackage{amsfonts}
\usepackage{amsmath}
\usepackage{amssymb}
\usepackage{multirow}
\usepackage{tabulary}

\theoremstyle{definition}

\journal{}

\begin{document}

\begin{frontmatter}

\title{Rank-based Heuristics for Optimizing the Execution of Product Data Models}


\author{Konstantinos Varvoutas\fnref{myaddress1}}
\author{Anastasios Gounaris\fnref{myaddress1}}
\author{Georgia Kougka\fnref{myaddress1}}
\author{Hajo A. Reijers\fnref{myaddress2}}

\address[myaddress1]{Aristotle University of Thessaloniki, Greece\\ \{kmvarvou,gounaria,georkoug\}@csd.auth.gr}
\address[myaddress2]{Universiteit Utrecht,The Netherlands\\ h.a.reijers@uu.nl}

\begin{abstract}
The Product Data Model (PDM) is an example of a data-centric approach to modelling information-intensive business processes, which offers flexibility and facilitates process optimization. Because the approach is declarative in nature, there may be multiple, alternative execution plans that can produce the desired end product. To generate such plans, several heuristics have been proposed in the literature. The contributions of this work are twofold: (i) we propose new heuristics that capitalize on established techniques for optimizing data-intensive  workflows in terms of execution time and cost and transfer them to business processes; and (ii) we extensively evaluate the existing solutions. Our results shed light on the merits of each heuristic and show that our new heuristics can yield significant benefits.
\end{abstract}


\end{frontmatter}

\section{Introduction}

Data-centric approaches have been emerging in the last two decades as an alternative to the more mainstream activity-oriented modelling approaches for business processes  \cite{evaluating,KunzleR11,HenriquesS10}. As stated in  \cite{evaluating} \emph{``The central idea behind data-centric approaches is that data objects/elements/artifacts can be used to enhance a process-oriented design or even to serve as the fundament for such a design"}. Apart from better modelling processing where data element production drives the whole process, data-centric approaches are reported to provide increased reusability and flexibility  in execution.

In this work, we focus on a particular data-centric modelling approach, namely a \textit{Product Data Model (PDM)-}oriented one. 
This approach is tailored to information-intensive processes and  is declarative in nature. 
The main driver for PDM, as also reported in \cite{evaluating}, is process optimization next to flexibility. A PDM instance focuses on describing what is needed in order to  deliver an information product rather than the exact way to achieve this goal. 
The declarative model is accompanied by a method to generate workflow designs, which is referred to as \emph{Product Based Workflow Support (PBWS)} \cite{pbws}. PBWS presents a set of heuristics for PDMs with the purpose of enhancing the performance on a case-by-case manner. PBWS improves upon a previous method, called
\emph{Product Based Workflow Design (PBWD)} \cite{ReijersLA03}, where the burden of  defining the sequence of actions rests with the workflow designer. While PBWD merely assists in this task through presenting the alternatives, PBWS defines the sequential order of task execution based on provided task metadata.

The problem we target in the present paper is to define the order of actions (a.k.a. operations) to execute in a process described by a PDM. Our optimization criteria include both execution cost and time duration. We aim to  improve the efficiency of  existing heuristic techniques in \cite{pbws} for defining such an ordering.
Heuristic solutions are intuitive in their rationale, easy to implement and are of low computational complexity. However, the existing solutions fail to benefit from established techniques in the areas of  optimization for workflows for data analytics and database query execution plans. Techniques from these domains adopt principled cost-based approaches \cite{Ibaraki84,KBZ86,KougkaGS18} and provide polynomial approximate solutions to NP-hard problems. Inspired by the existence of such techniques, in our heuristics, we suggest to consider  both the time and the cost of each operation in a PDM model and the probability of this operation to lead to an early termination of the process in a combined manner. This saves time and resources. More specifically, we make a twofold contribution:
\begin{enumerate}
	\item  We propose a new set of heuristics for choosing the next operation to be performed in a PDM for a specific case to optimize time duration and/or cost. Our proposals differ in the exact optimization criteria they take into account, the capability to establish whether an execution of an operation is meaningful, and in the manner they consider dependencies between the successful execution of operations.
	
	\item We perform an extensive experimental evaluation of the available heuristics and we show that our proposals yield benefits in terms of time, cost or a combination of both compared to previous heuristics, for the average case. 
	\end{enumerate}
	
Initial work to generate these contributions has appeared in \cite{VarvoutasG20}. The main extensions to our initial work include (i) an investigation of additional heuristics that differ in their rationale regarding the meaningfulness of operations and grouping of operation; (ii) an entirely new implementation  that addresses a number of limitations, such as not considering a PDM graph as a hypergraph\footnote{The implementation is publicly available from \url{https://github.com/kmvarvou/rank_based_heuristics} }; (iii) new experiments using additional, real-world PDM models and random parameterization, which provide a deeper insight into the performance of our proposal; and (iv) an investigation of the degree of approximation by the heuristics compared to the optimal solution.


The remainder of this paper is structured as follows. In Section \ref{sec:back}, we present the PDM underpinnings of the techniques evaluated. Section \ref{sec:rank} introduces  our proposal and Section \ref{subsec:implementation} discusses implementation. We  evaluate the candidate techniques in Section \ref{sec:eval}. Section \ref{sec:rw} deals with the related work and we conclude in Section \ref{sec:concl}, where we also briefly discuss limitations and future work.

\section{Background}
\label{sec:back}

In this section, we first introduce the main principles and notation regarding the PDM in Section \ref{sec:pdm} and we then discuss PDM execution in Section \ref{sec:pdm-exec}.

\subsection{The PDM}
\label{sec:pdm}

\begin{figure}[tb!]
	\centering
	\includegraphics[width=1.0\textwidth]{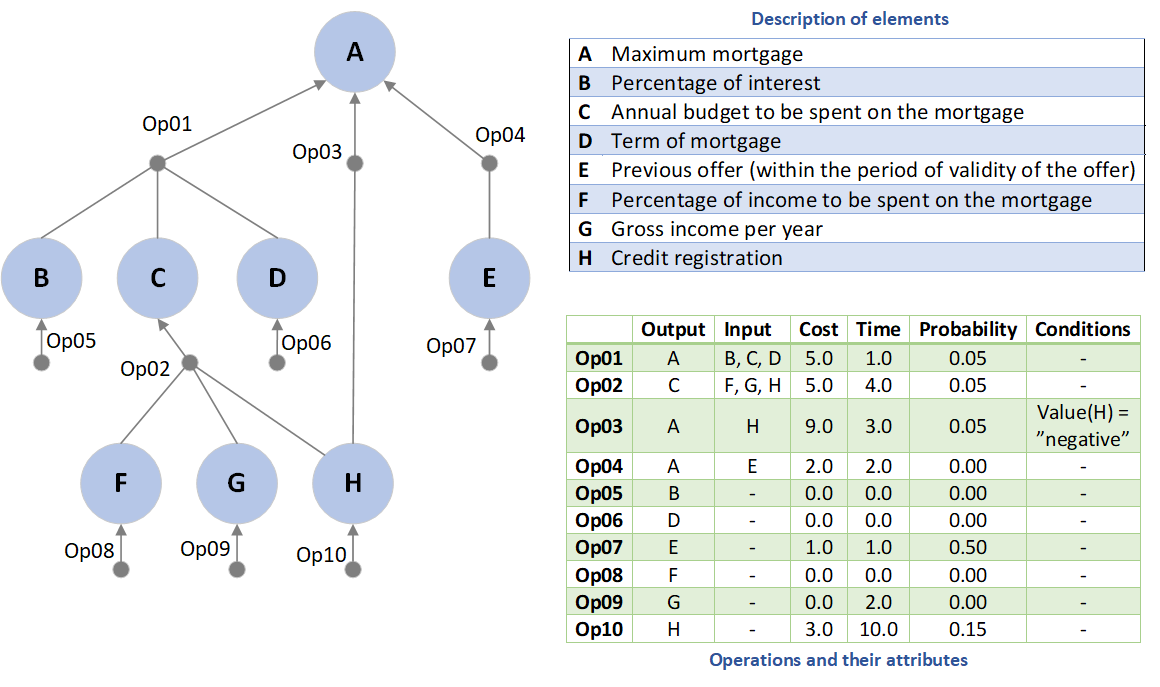}
	\caption{The PDM mortgage example from \cite{pbws}}
	\label{fig:pdm}
\end{figure}

A PDM is used to represent the structure of a \emph{workflow product} in a rooted graph-like manner, similar to a Bill of Material \cite{pbws,orlicky2003structuring}. 
A \emph{workflow} is a term used in many different ways; in our work,  we use this term to refer to a (potentially software system-supported) process  that passes information from one participant to another for action  emphasizing on how the tasks or activities of the process are carried out; this is in line with textbooks such as \cite{fund, Stiehl14}. The notion of a \emph{workflow product} covers the desired outcome of the process and the corresponding workflow \emph{without} providing a concrete execution plan on how this product is derived \cite{pbws}.
 
More specifically, PDMs describe the required elements for yielding the desired product in the root, where example (informational)  products include decision on whether to grant an approval to a specific admission request, approval of a mortgage application, and so on. In Figure \ref{fig:pdm}, we present such an example of a PDM for a classical mortgage application. The \emph{vertices} (or graph \emph{nodes}) in this structure correspond to data elements, which is the information that is processed in the workflow. The \emph{edges} correspond to operations to produce such data elements; operations are denoted as $Op$. Each operation has a set of quantitative metadata assigned to it, which typically differ between \emph{process instances} (or, equivalently, \emph{process cases}). 

The final product of the process is to determine the value of the root (or top or end product) vertex. To this end, values of multiple other data elements need to be determined from the leaves to the root, as specified by the edges of the graph. In essence, operations correspond to actions that are applied on the valued data elements to produce values for vertices downstream. Each operation may have zero or more input data elements. However, it produces the value for exactly one output data element at most; this occurs in the case that its execution is successful.

Let us assume that there are $n$ operations $Op_i,~ i \in \{1,\dots,n\}$  and $m$ data elements in a PDM.
An operation in a PDM is represented by a sextuple, denoted as $$Op_i=<Out_i, Input_i, cost_{i}, time_{i}, prob_{i}, cond_{i}>$$

$Out_i$ is the data element $j \in \{1,\dots,m\}$ that can be produced by $Op_i$. $Input_i$ contains the prerequisite data elements that serve as input to $Op_i$. 
To avoid cycles, which correspond to execution deadlocks, $Out_i$ cannot belong to the input set of any operations producing  the elements in its own input set as well. 
For example, in Figure \ref{fig:pdm}, the tuple  $<A, \{B, C, D\}, \dots>$ which corresponds to $Op_{01}$, describes the fact that $Op_{01}$ can be applied only if  data elements $B$, $C$ and $D$ are available and the execution of  $Op_{01}$  \emph{may} lead to the production of the value of the $A$ data element. Each operation is also accompanied by the following metadata, as defined in its sextuple; these metadata allow for cost-based decisions:

\begin{itemize}
	\item \emph{Cost ($cost_{i}$)}, which represents the cost associated with executing the operation $Op_i$, where $i \in \{1,....,n\}$.
	\item \emph{Time ($time_{i}$)}, which represents the time required for the complete execution of $Op_i$.
	\item \emph{Probability ($prob_{i}$)}, which represents the probability that $Op_i$ is executed \emph{unsuccessfully}, therefore not producing its output element.
	\item \emph{Conditions ($cond_{i}$)}, which represent requirements regarding the value of the input elements. These requirements must be met, if the operation $Op_i$ is to be executed, meaning that the existence of all input elements  is not sufficient; this is crucial in rendering PDMs useful in practice.
\end{itemize}   

Examples of these metadata are depicted in the table at the bottom right of Figure \ref{fig:pdm}. It is important to note that these metadata can refer to specific cases, e.g., in a loan application scenario, the cost may change according to the type of loan requested. 

In the case where an element has zero input elements, it is called a \emph{leaf element}. Commonly, leaf elements  are provided as input to the process (if their production is always guaranteed); in the figure, elements such as $B$, $F$ and $E$ are leaf elements. 
In addition, an operation is considered \emph{executable} when all of its input data elements have been produced, e.g., the operations that produce these elements have been successfully performed.
Finally, if an operation has multiple data element inputs, these inputs are bundled in the graph, e.g., as is the case for elements $B$, $C$ and $D$ for $Op_{01}$. 

In summary, the PDM describes the operations that can be performed to produce the root element, along with their inter-dependencies.  Not all operations need to be executed for the production of the root element. A data element may be produced through multiple operations, e.g., in the same example of Figure \ref{fig:pdm}, $A$ which is the root element, can be determined in three manners, namely, $Op_{01}$, $Op_{02}$ and $Op_{03}$. As a final note, we clarify that having a single root element does not prevent to model processes with multiple final outcomes through connecting all the corresponding final output elements to an artificial root element with zero-cost operations.

 \subsection{Execution of a PDM}
\label{sec:pdm-exec}

Product data models can be executed directly, without the need of a process model to support the execution of the process they represent \cite{Vanderfeesten09}. This is attributed to their flexible structure, which allows for the existence of multiple paths for the production of a particular element in a PDM. A \emph{path} is derived by a sub-graph of a PDM \cite{ComuzziV11} after performing a topological sorting. 
An ordered set, or equivalently, a sequence of operations that lead to the production of the root element is called a \emph{complete path}.

Each specific occurrence of the execution of a PDM is referred to as an \emph{instance} and is associated with a distinct \emph{execution path}, i.e.,  a sequence of operations chosen for this specific instance.
Overall, a PDM does not specify  how the end information product is created exactly. In this sense, it allows for multiple execution paths that produce the desired information product. 
In general,  the same data element in a PDM can be produced by multiple operations, each having a different set of input data elements, which allows for the existence of alternative ways of producing it. 
Therefore, PDMs usually feature alternative complete execution paths, each with a different cost and time duration.

The execution path of each instance may be determined either in an incremental step-by-step manner, where the next operation is determined only after the completion of the execution of the last operation, or before execution in full. As will be explained shortly, in this work, we follow the former approach, where the execution path is incrementally built. Given that the execution path described the order of operation execution, in our context this path coincides with the process \emph{workflow}. 

In order to evaluate the performance of our proposed approach, we  also need to introduce the notion of the \emph{optimal path}. As  explained above, a PDM usually features several alternative paths that produce its root element. An optimal path for a specific instance is a complete path that achieves the best possible performance in terms of a specific objective, such as execution cost, for that instance. 
A methodology to assess the extent of the approximation of a heuristic is to calculate the optimal path, for each of the evaluation’s instances, and use it as a reference point; i.e., detect all the complete paths and find the best performing one between them. According to the formal definition of the complete path, as presented by Comuzzi et al. \cite{ComuzziVW13}, the PDM mortgage example (Figure \ref{fig:pdm}) contains at least 3 complete paths; these paths may share some operations in common. These paths are presented in detail in Table \ref{tab:table-path} and, without loss of generality, let us assume that for the purpose of this work, we focus only on the cost metric only. According to the cost metric, the total cost of a path is the sum of the cost of the operations it contains. 

A key observation is that paths differing in the ordering of their constituent operations can be excluded from the minimal set of the complete paths because they do not differ in their total cost.\footnote{Here, we assume that an operation's cost is constant and does not differ depending on its position in the path.} That is, when we are interested only in the cost of the complete paths, these need not be interpreted in their typical sense in directed graphs but can be deemed as sets rather than lists/ sequences. For example, we do not include $Op_{08},Op_{09},Op_{10},Op_{02},Op_{05},Op_{06},Op_{01}$ as a separate path in the table, because it is equivalent to \emph{Path 1} in terms of performance. Ignoring the ordering of the operations does not have any impact on both the cost and the time duration of the corresponding execution path, due to the fact that operations are executed sequentially. 

\begin{table}[tb!]
	\begin{center}
		\begin{tabular}{|c | c | c|} 
			\hline
			Path & Operations & Total Cost\\ [0.5ex] 
			\hline
			\emph{Path 1} & $Op_{07}, Op_{04}$ & 3\\
			\hline
			\emph{Path 2} & $Op_{10}, Op_{03}$ & 12\\
			\hline
			\emph{Path 3} & $Op_{10}, Op_{09}, Op_{08}, Op_{02}, Op_{05}, Op_{06}, Op_{01}$ & 13\\ 
			\hline
		\end{tabular}
		\caption{\label{tab:table-path}Minimal set of complete Paths of the mortgage example PDM in Figure \ref{fig:pdm}.}
	\end{center}
\end{table}

There are two additional important remarks when determining the optimal path of a PDM instance. The first one  relates to the specific attribute values that are specific to this instance, which implies that the optimal path may be different across multiple instances of the same PDM process. This typically occurs when the time and durations of the operations fall into a certain range, i.e., they are not the same for all instances.
But even if the cost and the duration of an operation is fixed, when the probability of failure is not 0 or 1, the execution of an operation is not deterministic  in a manner that  may impact on the optimality of a path.
The second remark extends the point above and relates to the \emph{feasibility} of the optimal path, which may alter during the instance's execution, due to a failed operation. For example, \emph{Path 1} is the optimal path of the mortgage PDM in terms of cost, since it has a total cost of 3. However, the feasibility of \emph{Path 1} depends on the successful outcome of its operations. If $Op_{04}$ fails, then \emph{Path 1} is no longer feasible, and another path becomes the optimal one. In this case, the  feasible path with the lowest cost for such an instance is \emph{Path 2}. Overall, the optimal path for a specific instance  reflects this instance's specif details in terms of element availability, and therefore path feasibility. 

\section{Proposal}
\label{sec:rank}

The optimization problem we deal with is: \emph{which paths of operations to choose for a specific PDM instance in order to optimize user-defined quantitative objectives of cost and time?}

We address this problem in a case-by-case manner, i.e., if different instances come with different metadata, such as $cost_i$, $time_i$ and $prob_i$ values, different sequences of operations may be devised. Moreover, the execution path is built incrementally taking into account whether the last operation chosen was executed successfully or not.

First, in Section \ref{sec:approach} we describe the simpler form of our solutions, which rely on the \emph{rank} metric. Then, in Sections \ref{subsec:meaning} and \ref{sec:group}, we extend our solutions to allow for more judicious decisions.

\subsection{Main solution}
\label{sec:approach}

There are two high-level strategies for the calculation of an optimal execution path corresponding to a PDM instance, namely a global and a local one. A global strategy considers the effect of each decision on future steps. It takes into account the complete set of alternative paths that produce the end product to optimize the execution performance of each case. Instead, a local strategy adopts a step-by-step approach, meaning that, at each step, it examines the set of executable operations and chooses the best one, according to a particular metric, e.g. cost of execution. As explained in \cite{pbws}, a global strategy does not scale as the problem is NP-hard. For this reason, in this work, we exclusively deal with low-polynomial local strategy heuristics. For small-scale models, we also derive the optimal solution using exhaustive techniques. This is done to provide insights into how closely our heuristics approximate the optimal solutions.

Our approach to constructing execution paths relies on treating operations in a manner similar to knock-out business process activities, and the manner their optimal ordering is decided. This also bears similarities to the way data analytics operators and database selections and joins are ordered \cite{reeng,transfer,KougkaGS18,Ibaraki84,KBZ86}.
A knock-out activity is an activity whose execution leads directly to the completion of the process \cite{reeng}. Typically,  operators with   low cost and low selectivity (i.e., keeping as few tokens or data tuples as possible) are ordered earlier. The metaphor in our case is that reaching the root element is similar to knock-out activities eliminating tokens in the sense that the path is completed.
For example, the execution of $Op_{03}$ in Figure \ref{fig:pdm}, which produces the root element $A$ is a knock-out operation. Then, the optimal ordering needs to take into account the probability of an operation to produce the end element and the corresponding cost or execution duration. An important point is that an operation may also  contribute to the production of the root element indirectly, e.g., through  a sequence of operations starting with that operation and ending to the root element. 

More specifically, at each step where the next operation is selected, we consider all the executable operations (i.e., those with all input elements available) and we choose the one with the highest \emph{rank} value. The rank value of an operation $Op_i$ is defined as follows:
\begin{equation}
rank(Op_i) = \frac{\prod_{Op' \in \pi(Op_i)}1-prob_{Op'}}{\sum_{Op' \in \pi(Op_i)}Cost_{Op'}}  
\label{eq:rank}
\end{equation}

where $\pi(Op_i)$  is the path from $Op_i$ (including) to an operation directly producing the root. 
The function essentially defines that the rank value is (i) proportional to the probability of $\pi(Op_i)$ to execute without any operation failures and (ii) inversely proportional to the sum of the costs of the operations in the path.

\begin{table}[tb!]
	\begin{center}
		\begin{tabular}{|c | c | c|} 
			\hline
			operation & path & rank value\\ 
			\hline \hline
			$Op_{02}$ & $Op_{02}, Op_{01}$ & 0.09025\\
			\hline
			$Op_{03}$ & $Op_{03}$ & 0.105556\\
			\hline
		\end{tabular}
		\caption{\label{tab:table-rank-example}The executable operations along with their rank values in the example.}
	\end{center}
\end{table}


\begin{figure}[tb!]
	\centering
	\includegraphics[width=0.495\textwidth]{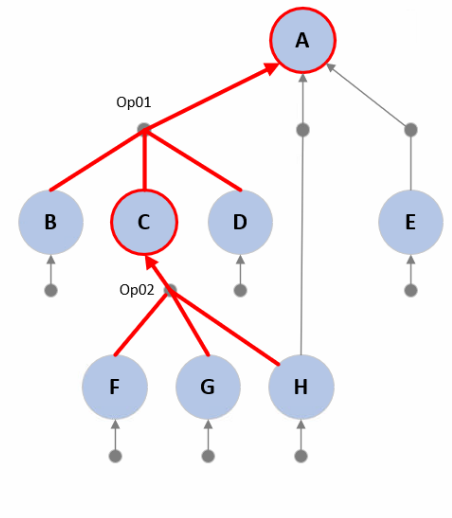}
	\includegraphics[width=0.495\textwidth]{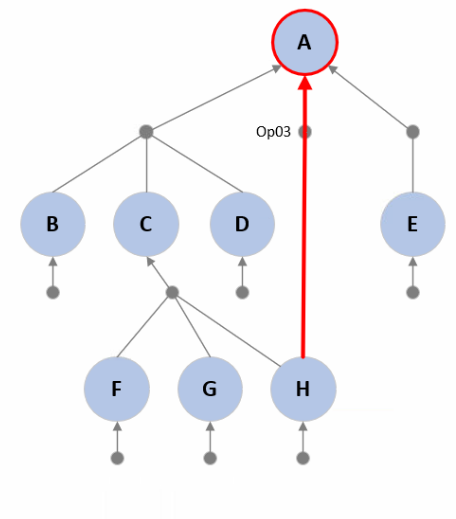}
	\caption{The two execution paths that lead to the production of the root element and correspond to operations $Op_{02}$ and $Op_{03}$ respectively.}
	\label{fig:pdm_paths_example}
\end{figure}

\emph{Example:} In the example in Figure \ref{fig:pdm}, let us assume a state in which all leaf elements have been produced already except  element $E$, for which $Op_{07}$ was not executed successfully.  Thus, in the next step, there are  two executable operations, namely $Op_{02}$ and $Op_{03}$. Table \ref{tab:table-rank-example} presents the alternatives, which are also highlighted in Figure \ref{fig:pdm_paths_example}. Based on its attributes, $Op_{02}$ has the following ranking value: $rank(Op_{02}) = 0.9025/10 = 0.09025$.  This is because, while $Op_{02}$ may not lead to a process termination directly, we consider $Op_{02}$ as part of a path that leads indirectly to the root, that is the path: ($Op_{02}, Op_{01}$). Therefore, we use as probability of this path,
the probability of \emph{success} of the operations in the whole path as denoted in the nominator of the rank function, which is $(1-prob_{Op_{02}})*(1- prob_{Op_{01}}) = 0.95*0.95 = 0.9025$ and as cost,  the aggregate cost of the whole path, which is $Cost_{Op_{01}} + Cost_{Op_{02}} = 5+5 = 10$. On the other hand, 
$rank(Op_{03}) = 0.95/9 = 0.105556$. The probability 0.95, in the nominator, is the probability of the \emph{successful} execution of $Op_{03}$ because it is the successful execution of $Op_{03}$ that produces the root element $A$ and therefore, completes the workflow execution. Based on these values, $Op_{03}$ is selected for execution. 

Note that 
we have not discussed how we treat the case where there are multiple paths starting from the same operation yet. The relevant explanations and clarifications are provided in the next section.

In summary, the proposed rank-based heuristic chooses, at each step, the operation, which is ready to be executed and has the highest rank value.
This heuristic comes in three variants. The first one uses the formula above. The other two modify the denominator  in the rank formula and employ (i) the sum of the operation duration times and (ii) the sum of both the normalized operation cost and times, respectively\footnote{In our preliminary work \cite{VarvoutasG20}, the latter flavor employed the product of the cost and times; we have modified this variant as the sum of these attributes exhibits higher performance.}. As such, the other two variants focus more on time duration and a combination of both metrics, respectively.

Building upon this three-flavored heuristic, we  introduce an additional multi-variant heuristic next. 
The extended rank-based heuristic employs the same rank formula, but it is different in the fact that it relies on the notion of \emph{meaningless operations}. 


\subsection{Meaningful and meaningless operations}
\label{subsec:meaning}

In this section, we introduce the notion of \emph{meaningless} operations, which forms the basis for our extended rank-based heuristics. These heuristics do not rely on the rank value solely to decide the next operation.

Remember that the execution of a PDM and the corresponding execution path construction is  dynamic  and differs across multiple instances of the same model even if the same heuristic is employed. Such variability is dependent on whether an operation chosen for execution is successful or not.  The failure of a particular operation may render any complete path containing this operation unfeasible and any operations belonging to unfeasible paths exclusively have no reason to execute. For example, consider the PDM of Figure \ref{fig:pdm}. Elements $F$, $G$ and $H$ are all inputs to $Op_{02}$. We assume an execution instance where the first operation selected for execution is $Op_{08}$, and it is executed unsuccessfully. Consequently, regarding $Op_{09}$, which produces element $G$, there is no reason to execute, despite being executable. Due to the failure of $Op_{08}$, $Op_{02}$ cannot be executed, and therefore the execution of $Op_{09}$ would not lead to the production of the root element in any case. On the contrary, $Op_{10}$, which produces element $H$ can still be executed due to its contribution to the possible execution of $Op_{03}$. Operations, such as $Op_{02}$ will be referred to as meaningless operations. Below, we present the formal definition. 


More specifically, at any state while constructing an execution path referring to a PDM instance,  
if an operation $Op_i,~i \in \{1,\dots,n\} $, which produces element \emph{A} has already been executed unsuccessfully, an operation $Op_j, ~j \neq i$ and $i \in \{1,\dots,n\}$ that has not already executed, which produces element \emph{B}, is a \emph{meaningless} operation if and only if both conditions below hold: 
\begin{enumerate}
    \item There exists at least one operation that has not executed yet,  such that it features both elements \emph{A} and \emph{B} as its inputs.
    \item There exists no operation that has not executed yet such that it features element \emph{B} as its input without having element \emph{A} as its input.
\end{enumerate}

The  rationale behind the conditions above is as follows. First, we will reason that if any of the two statements does not hold, then we cannot discard the corresponding operation. If there is no operation that has both  \emph{A} and \emph{B} in its input, then the fact that \emph{A} cannot be produced does not block the path from \emph{OpB} to the root. The second statement says that, for $Op_j$ to be meaningless,  for every operation that \emph{B} appears on its input list, \emph{A} has to be on the same list as well. If this does not hold, then there is at least one path connecting $Op_j$ to the root that remains feasible. Finally, we have to also prove that when both conditions hold, then executing $Op_j$ does not contribute to the production of the root element. This is true, since all paths to the root are blocked due to the co-presence of \emph{A} and \emph{B} in the input list of an operation in such a path. 


The notion of the meaningful operations is leveraged in two ways in the extended rank-based heuristic. First, only operations deemed  non-meaningless are considered when choosing the next operation to be executed. In this manner, the cost of operations, whose execution doesn't contribute anything to the production of the root element is saved. Therefore, the consideration of only meaningful operations constitutes an effective pruning of sub-optimal choices. Furthermore, it is also used to detect interdependent data elements. This notion is explained in detail, in the next subsection, which provides the complete view of the extended heuristic.

\subsection{Grouping Interdependent Elements}
\label{sec:group}
A main idea behind the extended rank-based heuristic is, during deciding the next operation to execute, to group elements, whose impact on the production of the root element relies on the availability of each other. For example, the usability of element $G$ depends upon the availability of element $F$ (and vice versa) in order to have any impact on the production of the root element, via $Op_{02}$. By contrast, element $H$’s impact is independent of the availability of any other element because it can trigger an operation leading to the root element, either directly or indirectly, without any further prerequisites. Based on this observation, the group-based heuristic assigns a rank value to 
the element $G$ taking into account its interdependence with element $F$ (and vice versa). 

In this heuristic, elements are grouped together whenever they are deemed to be interdependent. More formally, two elements are interdependent when they appear as inputs in exactly the same set of operations, like elements $F$ and $G$ mentioned above. In practice, to detect which elements should be grouped together, we rely on the notion of meaningless operations, which has been introduced before.  For each element $j$, $j \in \{1,\dots,m\}$, we determine which elements are dependent upon its availability. That is, for each element $j$,  we determine the set of elements whose production would be \emph{meaningless} if this particular element $j$ could not be produced. For every element in the latter set, its distance from the root is increased by the production cost of element $j$. For example, a data element dependent on two other elements would have its distance from the root increased by the sum of the production costs of these two elements.

\section{Implementation}
\label{subsec:implementation}
To apply our heuristics described previously, which involve paths from arbitrary vertices to the root element,  we need a data model that represents a PDM as a directed graph, on which traditional graph theoretical algorithms, such as shortest paths can be computed.  
In this section, we first discuss the transformation of PDMs to directed graphs in Section \ref{subsec:repres}.  
Then, we continue with  explaining the exact manner we detect paths and meaningfulness of operations in Section \ref{subsec:rank-path}  and we close discussing the data structures and complexities of the underlying algorithms in Section \ref{subsec:complexities}.

\subsection{Representation of PDMs}
\label{subsec:repres}

\begin{figure}[tb!]
	\centering
	\includegraphics[width=0.47\textwidth]{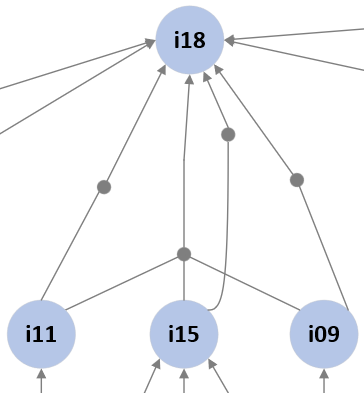}
	\includegraphics[width=0.47\textwidth]{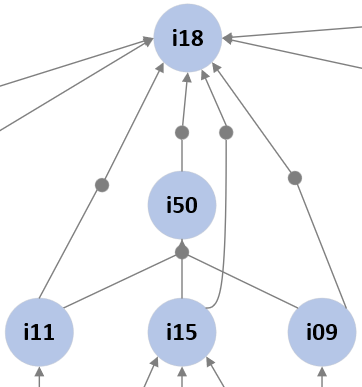}
	\caption{A part of an original PDM model (left) and the insertion of a dummy vertex (right).}
	\label{fig:pbwddummy}
\end{figure}

A PDM is a special type of directed hypergraph \cite{Comuzzi17}, which generalize  directed graphs \cite{AusielloFF01}. More specifically, the edge of a hypergraph, called hyperedge, may have multiple source and destination vertices. Moreover, in a PDM, there can be more than one edge connecting two vertices, similar to a multigraph. For example, on the left pert of Figure \ref{fig:pbwddummy}, we focus on an excerpt of a  social insurance PDM, shown in full   in Figure \ref{fig:pdm2} and used in our evaluation. In this part of the PDM, the elements  $i11$ and $i18$ are connected directly by two different operations. The same applies to elements $i15$ and $i18$. 
To transform the PDM graph into a directed graph representation,  
we add a set of \emph{dummy}, i.e. artificial, vertices in addition to the original data elements.  These vertices are added with the view to enforcing that the number of direct edges between every pair of element vertices is at most one. 

In Figure \ref{fig:pbwddummy}(right), we illustrate the addition of a dummy vertex in the aforementioned case of the social insurance PDM of Figure \ref{fig:pdm2}. In the transformed model, a new element $i50$ has been added. In addition to the new element, a new artificial operation, which takes $i50$ as an input and produces element $i18$, has also been added. Such artificial operations do not modify any semantics  of the original model. To ensure that the addition of dummy elements has no impact on the execution behavior of the PDM, the  attributes of the corresponding artificial operations  are set as follows: the execution cost, the execution time and the probability of failure are all set to 0. 

\subsection{Computing the rank value}
\label{subsec:rank-path}
In the rank-based solutions, we have used the notion of  $\pi(Op_i)$, which  is a path from $Op_i$ to an operation producing the root element. However, in general, there is not a single such path, which implies that there is no deterministic manner to compute  $rank(Op_i)$ and more importantly, costly paths to be selected instead of less expensive ones.

To address this issue, that is to   provide an efficient and deterministic manner to compute $rank(Op_i)$ in any PDM, we adopt the following methodology.
We use  the probability of failure of data element production, exactly as provided in the metadata table to assign weights to the graph edges. Then,  we choose  the path for which the sum of these probabilities is the smallest one as the representative path of the operator $Op_i$, i.e., we reduce the problem to a shortest path one for which simple algorithms, such as Dijkstra's, can be employed. In addition, 
the path in the denominator need not be the same as in the nominator and can be detected applying directly a shortest path algorithm after assigning to each graph edge the corresponding operation cost (or time duration) as its weight.

\subsection{Data Structures and  Complexities}
\label{subsec:complexities}

There are four auxiliary data structures involved: (i) a dictionary structure, e.g., implemented as a 
\texttt{HashMap} that stores the operation metadata for each operation; (ii) a list that contains the operations ready to be executed; (iii) a list that contains the elements already produced and (iv) a dictionary that connects the output element of an operation to the operations that it serves as their input. The latter is required to identify meaningless operations. 

In the beginning, the two lists are initialized through a simple traversal of the operation metadata. The complexity of this is  $O(nm)$ where $n$ is the number of operations and $m$ the number of the vertices in the graph. The $n$ and $m$ values include also the dummy zero-cost operations and vertices. 

Then, the heuristics repeat as many steps as the number of operations in the execution path, which is up to $n$. In each step, the next operation is chosen after examining $n$ operations at most. Also, after the selection and execution of the next operation,  $O(m)$ checks are performed to update the list of eligible operations.\footnote{More efficient implementation can be devised, but  since PDMs are not particularly large containing a few dozens of operations and elements, employing more sophisticated data structures does not pay off.}
Finding the shortest path from a root element is at most $O(n logm)$ using an algorithm such as Dijkstra. Identifying meaningless operations after each failed operation execution is $O(nm)$ using the second dictionary.
Thus the overall complexity of the rank-based techniques is $O(n^2(n+m)logm)$, which is $O(n^3logm)$, since $n \ge m$. 




\section{Evaluation}
\label{sec:eval}

We begin this section with the description of the experimental setting and the competitors in Section \ref{subsec:competitors}. The main results are presented in Sections \ref{sssec:fs}-\ref{sssec:ss}.  Section \ref{sssec:other} investigates the impact of early termination and the efficiency of our proposals. In Section \ref{sssec:additional}, we discuss additional issues, such as comparison against the results in our earlier work. Finally, Section \ref{sssec:opt} discusses issues regarding the deviation of optimal solutions, when it is feasible to compute them.

\begin{figure}[tb!]
	\centering
	\includegraphics[width=0.995\textwidth]{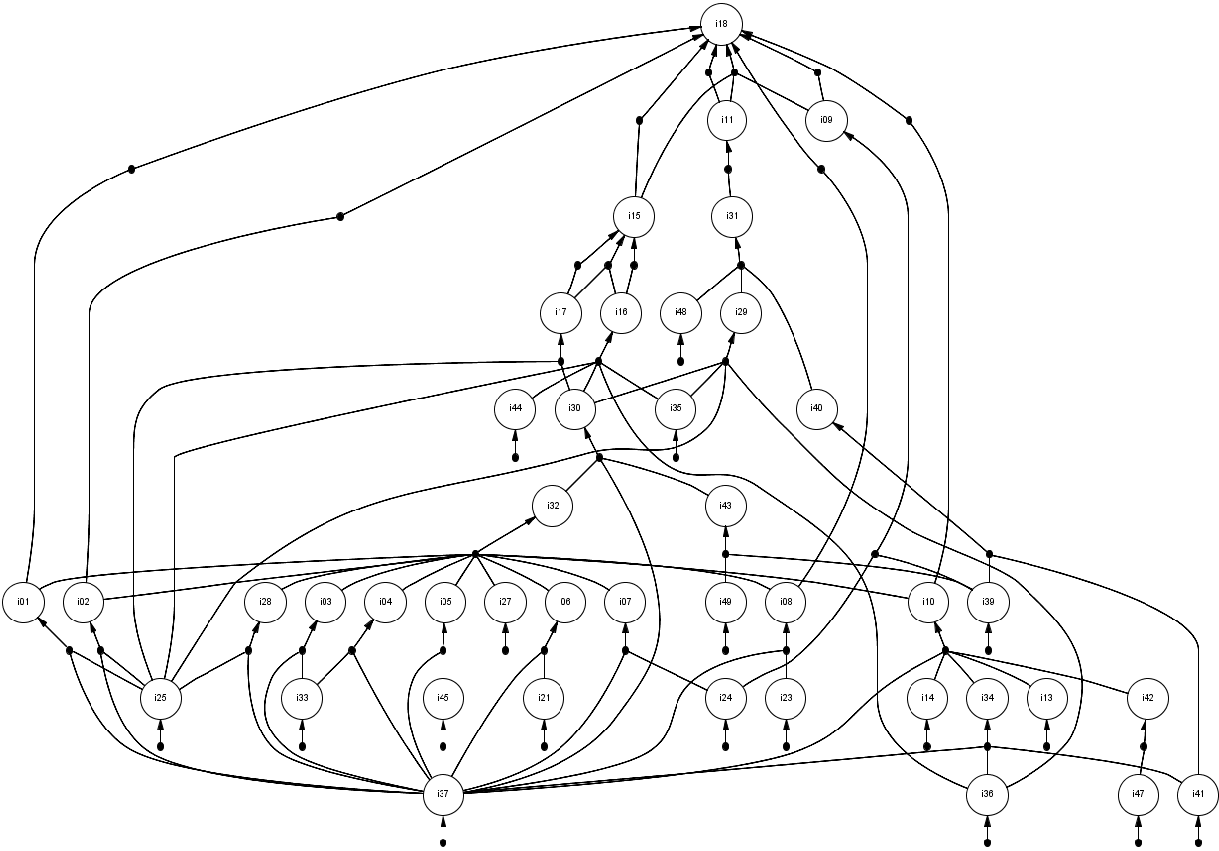}
	\caption{The PDM social insurance example from \cite{ReijersLA03}}
	\label{fig:pdm2}
\end{figure}

\subsection{Competitors, test PDMs and experimental settings}
\label{subsec:competitors}

Our competitors are the local strategies used by the PBWS method in \cite{pbws} comprise the following heuristics, which are referred to as \emph{existing heuristics}:
\begin{enumerate}
	\item \emph{Random}: the next operation is randomly selected from the set of executable operations.
	\item \emph{Lowest Cost}: the operation with the lowest cost is selected to be executed next.
	\item \emph{Shortest Time}: the operation with the shortest time is selected to be executed next.
	\item  \emph{Lowest Failure Probability}: the operation with the lowest probability of not being executed successfully is selected.
	\item \emph{Shortest Distance to Root Element}: the operation with the shortest distance to the root element (measured in the total number of operations) is selected.
	\item \emph{Shortest Remaining Process Time}: the operation with the shortest remaining processing time (measured as the sum of the processing times of the operations on the path to the root element) is selected.
	\item \emph{Shortest Remaining Cost}: the operation with the shortest remaining cost (measured as the sum of the costs of the operations on the path to the root element) is selected.	 
\end{enumerate}

Note that the last heuristic is not explicitly mentioned in \cite{pbws} but it is trivial to devise it.

\begin{figure}[tb!]
	\centering
	\includegraphics[width=0.8\textwidth]{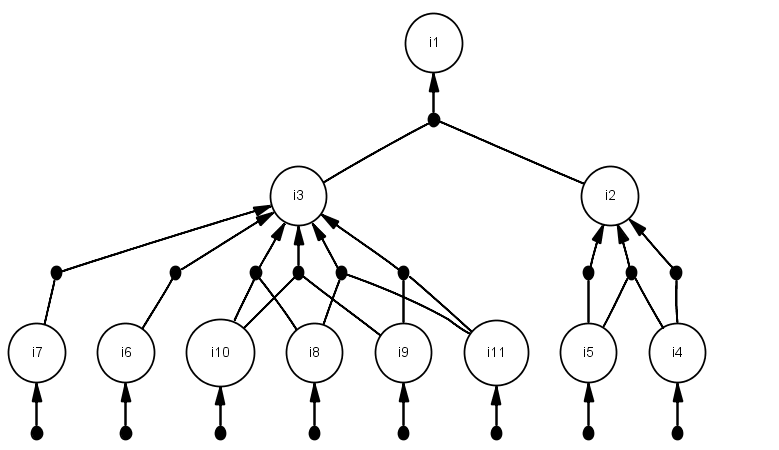}
	\caption{The PDM monitoring example from \cite{ComuzziV11}}
	\label{fig:pdm3}
	\end{figure}

We evaluate our proposed approach using three PDMs found in the literature. The first one is the mortgage example that was presented in Section \ref{sec:back}, while the second one, shown in Figure \ref{fig:pdm2}, represents a larger process from a social insurance company \cite{ReijersLA03}. These two PDMs constitute the evaluation set that was also used in our previous work \cite{VarvoutasG20}. To provide a more thorough evaluation, we expanded this set with an additional, third PDM, as shown in Figure \ref{fig:pdm3}, which represents a monitoring process from \cite{ComuzziV11}.

To evaluate the 7 existing heuristics and our 6 rank-based flavors, we created two experimental settings. Each consists of 10K random cases for each of the aforementioned PDMs. The 6 rank-based flavors are the three ones, as described in the main part of Section \ref{sec:rank} along with their 3 extended counterparts that perform grouping and leverage the detection of meaningless operations.

In the first setting, we rely on the original metadata included with each PDM, as these are reported in their respective publications (see the Appendix).  Based on these metadata, we generate, for each operation's cost, time, and failure probability attributes, values according to a $N(\mu, \sigma^{2})$ (normal distribution). Here $\mu$ is the original value of that particular operation and $\sigma^{2}$ is the variance of the data. 
For each of the three PDMs the mean value of the cost and time attributes is derived from the original metadata. In the first and the third PDM, the standard deviation was set as a fraction of these mean values, 0.5X and 0.33X, respectively (otherwise, the variance would be very large, something that we explore in the second setting). In the second PDM, the variance was set according to the original metadata accompanying the model. 

In the second setting, and with a view to evaluating the behavior of the PDM operation ordering solutions in a wider range of metadata values, we also rely on synthetic values for the attributes of the PDMs' operations. However, in this case,  the cost and time attributes were assigned (integer) values drawn randomly from a uniform distribution in the [0,10] range, i.e., these attributes may differ up to an order of magnitude. The probability of failure was assigned values in the [0.0, 1.0] range, i.e., we consider the complete range from guaranteed success to  certain failure. 

In both settings, we do not explicitly consider the probability of meeting conditions, since this probability affects the production of the output data element and it can be deemed as covered by the failure probability.  Finally, in order to conduct a fair comparison between the alternatives that encapsulate the detection of meaningless operations and the other ones, we initially report results only regarding the instances, in which the  root element could be successfully produced.
\subsection{First Setting: operation attributes follow a Gaussian distribution}
\label{sssec:fs}

\begin{table}[tb!]

	\begin{center}
	\resizebox{0.995\textwidth}{!}{
		\begin{tabular}{ccccccccc} 
			\hline
			\bf{Heuristic} &  \bf{Aggregate }  & \bf{Aggregate} & \bf{PDM1} & \bf{PDM1} & \bf{PDM2} & \bf{PDM2} & \bf{PDM3}  & \bf{PDM3}\\  
			 &  \bf{Cost}  & \bf{Time} & \bf{Cost} & \bf{Time} & \bf{Cost} & \bf{Time} & \bf{Cost} & \bf{Time}
			\\
			\hline
			\multicolumn{9}{c}{Existing heuristics} \\ \hline
		    Random &  3.59 & 3.85 & 3.31 & 3.73 & 5.53 & 5.48 & 3.26 & 3.45\\
			\hline
			Lowest Cost & 1.93 & 3.12 & 1.82 & 3.10 & 3.07 & 4.53 & 1.67 & 2.70\\
			\hline
			Shortest Time  & 2.58 & 1.85 & 2.11 & 1.65 & 4.57 & 3.08 & 2.41 & 1.66\\ 
			\hline
			Lowest Failure Probability &  3.50 & 3.61 & 4.61 & 4.91 & 2.19 & 2.38 &   {\bf 1.36} &  {\bf 1.34}\\ 
			\hline
			Root Distance &  1.98 & 2.00 & 1.69 & 1.53 & 2.20 & 2.23 & 2.19 & 2.38\\ 
			\hline
			Remaining Cost &  {\bf 1.46} & 1.78 &  {\bf 1.16} &  {\bf 1.21}  &   {\bf 1.55}  & 2.28 & 1.72 & 2.19\\ 
			\hline
			Remaining Time &  1.93 &  {\bf 1.59} & 1.67 & 1.36 & 2.31 &   {\bf 1.57} & 2.07 & 1.82\\ 
			\hline
			\multicolumn{9}{c}{Our proposals} \\ \hline
		    Rank-Cost &  1.36 & 1.91 & 1.32 & 1.60 & 1.64 & 2.45 & 1.32 & 2.04 \\ 
			\hline
			Rank-Time &  2.02 & 1.49 & 1.97 & 1.59 & 2.49 & 1.65 & 1.92 & 1.33\\ 
			\hline
			Rank-Combo &  1.55 & 1.52 & 1.58 & 1.51 & 1.72 & 1.72 & 1.46 & 1.45\\ 
			\hline
			Rank-Extended-Cost &  {\bf 1.22} & 1.67 &  {\bf 1.03} & 1.17 &  {\bf 1.48} & 2.04 &  {\bf 1.32} & 2.04\\ 
			\hline
			Rank-Extended-Time &  1.59 &  {\bf 1.23} & 1.10 &  {\bf 1.04} & 2.05 &  {\bf 1.48}  & 1.92 &  {\bf 1.33}\\ 
			\hline
			Rank-Extended-Combo &  1.33 & 1.31 & 1.11 & 1.08 & 1.59 & 1.59 & 1.46 & 1.45\\ 
			\hline
		\end{tabular}}
		\caption{Normalized performance of each heuristic for the first setting (the lowest the better).}
		\label{tab:setting1}
	\end{center}
	
\end{table}

We begin our discussion of the first experimental setting, where the operation costs follow a gaussian distribution with mean values based on real-world evidence, with the summary results in Table \ref{tab:setting1}. The values are normalized, so that, in each experiment instance, the value 1 corresponds to the most efficient cost or time time, respectively.
In the table, the best performing existing heuristic and the best performing variant of our proposals are marked in bold. There are two main observations that can be drawn:
\begin{enumerate}
    \item Our proposal consistently outperforms existing heuristics and more importantly, there are variants that consistently outperform all other variants: when the optimization objective is cost, our extended rank-based variant aiming at cost yield the lowest cost both on the average case and when checking each PDM case study separately; similarly, when the optimization objective is time, the best performing variant is the extended variant aiming at time explicitly. On average, the improvements over the best heuristic  overall are 16.5\% lower cost (when aiming at cost) and  22.6\% shorter time (when aiming at time).
    
    \item There is no dominant solution among the existing heuristics. As shown in the upper part of the table, a different existing heuristic is the best performing one when considering all PDMS and when considering each PDM separately. However, the best performing heuristics outperform the non-extended rank-based proposals.
\end{enumerate}

\begin{figure}[tb!]
	\centering
	\includegraphics[width=0.495\textwidth]{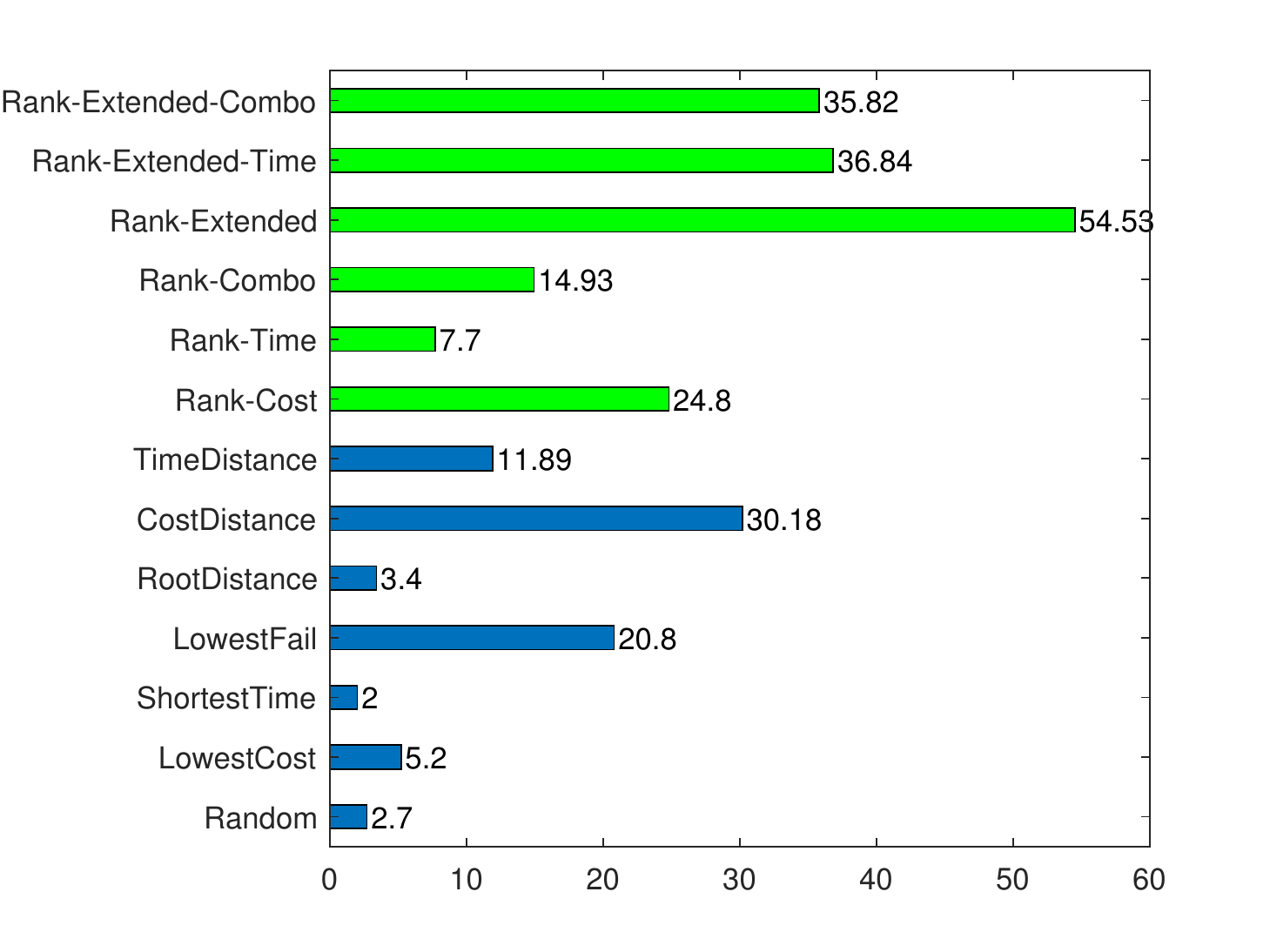}
	\includegraphics[width=0.495\textwidth]{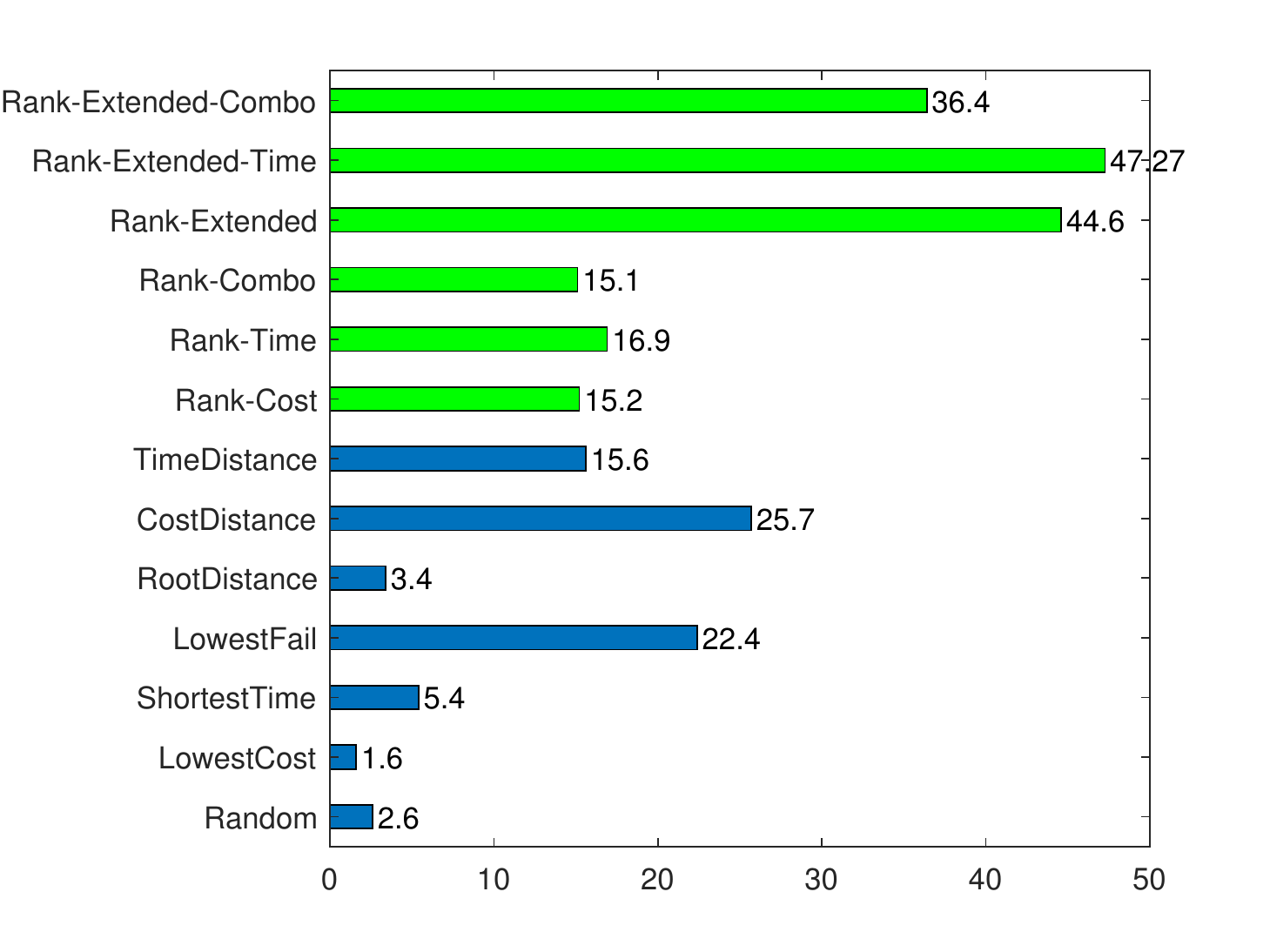}
	\caption{The percentage of cases where each heuristic achieved the best result in terms of cost (left) and time (right) in the first setting of  the evaluation.}
	\label{fig:set1_cases}
\end{figure}

Next, we move to Figure \ref{fig:set1_cases}, where we see the percentage of cases, where each solution achieved the best performance. The values do not sum up to 100\% because, in a random instance, more than one solution may have yielded the best performance and/or some instances do not lead to the production of the root element, as already explained. As shown, our extended rank heuristics yield better execution plans more frequently than the other variants. 

\subsection{Second Setting: operation attributes follow a uniform distribution}
\label{sssec:ss}

\begin{table}[tb!]

	\begin{center}
	\resizebox{\textwidth}{!}{
	\begin{tabular}{ccccccccc}
			\hline
			\bf{Heuristic} &  \bf{Aggregate }  & \bf{Aggregate} & \bf{PDM1} & \bf{PDM1} & \bf{PDM2} & \bf{PDM2} & \bf{PDM3}  & \bf{PDM3}\\  
			 &  \bf{Cost}  & \bf{Time} & \bf{Cost} & \bf{Time} & \bf{Cost} & \bf{Time} & \bf{Cost} & \bf{Time}
			\\
			\hline
			\multicolumn{9}{c}{Existing heuristics} \\ \hline
		    Random &  4.11 & 4.08 & 3.97 & 4.01 & 5.68 & 5.48 & 2.11 & 2.13\\
			\hline
			Lowest Cost & 2.39 & 3.66 & 2.02 & 3.62 & 3.61 & 4.86 & 1.65 & 1.94\\
			\hline
			Shortest Time  & 3.68 & 2.39 & 3.57 & 2.00 & 5.08 & 3.66 & 1.88 & 1.61\\ 
			\hline
			Lowest Failure Probability &  2.89 & 2.80 & 2.82 & 2.73 & 3.84 & 3.72 & 1.63 & 1.63\\ 
			\hline
			Root Distance & 2.53 & 2.57 & 3.02 & 3.04 & 2.11 & 2.21 & 1.68 & 1.71\\ 
			\hline
			Remaining Cost  &  {\bf 1.55} & 2.39 & {\bf 1.49} & 2.57 & {\bf 1.75} & 2.48 & {\bf 1.40} & 1.68\\ 
			\hline
			Remaining Time &  2.40 & {\bf 1.56} & 2.54 & {\bf 1.48} & 2.59 & {\bf 1.81} & 1.68 & {\bf 1.41}\\ 
						\hline
			\multicolumn{9}{c}{Our proposals} \\ 			\hline
		    Rank-Cost &  1.32 & 1.64 & 1.31 & 1.75 & 1.36 & 1.61 & 1.27 & 1.37\\ 
			\hline
			Rank-Time & 1.67 & 1.31 & 1.79 & 1.29 & 1.64 & 1.37 & 1.39 & 1.27\\ 
			\hline
			Rank-Combo &  1.44 & 1.42 & 1.47 & 1.45 & 1.44 & 1.43 & 1.31 & 1.30\\ 
			\hline
			Rank-Extended-Cost & {\bf 1.20} & 1.41 & {\bf 1.13} & 1.38 & {\bf 1.29} & 1.49 & {\bf 1.27} & 1.37\\ 
			\hline
			Rank-Extended-Time & 1.44 & {\bf 1.20} & 1.42 &  {\bf 1.13} & 1.51 & {\bf 1.29} & 1.39 & {\bf 1.27}\\ 
			\hline
			Rank-Extended-Combo &  1.28 & 1.27 & 1.13 & 1.24 & 1.34 & 1.33 & 1.31 & 1.30\\ 
			\hline
		\end{tabular}}
		\caption{Normalized performance of each heuristic for the second setting (the lowest the better).}
		\label{tab:setting2}
	\end{center}
	
\end{table}

In the second setting, the metadata values follow a uniform distribution. The advantage of this setting is that it covers a broader range of values; the disadvantage is that it may depart from real-world settings. Nevertheless, the outcomes of these experiments further strengthen the evidence that our rank-based variants advance the state-of-the-art. More specifically, the main observation drawn from Table \ref{tab:setting2} is twofold:
\begin{enumerate}
    \item Our extended rank based variant continues to consistently outperform existing heuristics. The average improvements over all random instances are  lower cost by 22.6\%  and shorter time by 23.1\%.
    \item In this setting, there is a clear winner among the existing heuristics, namely \emph{Shortest Remaining Cost} for the cost objective and \emph{Shortest Remaining Process Time} for the time objective. However, these heuristics cannot outperform event the non-extended rank-based heuristics, an observation already made in \cite{VarvoutasG20}.
\end{enumerate}

\begin{figure}[tb!]
	\centering
	\includegraphics[width=0.495\textwidth]{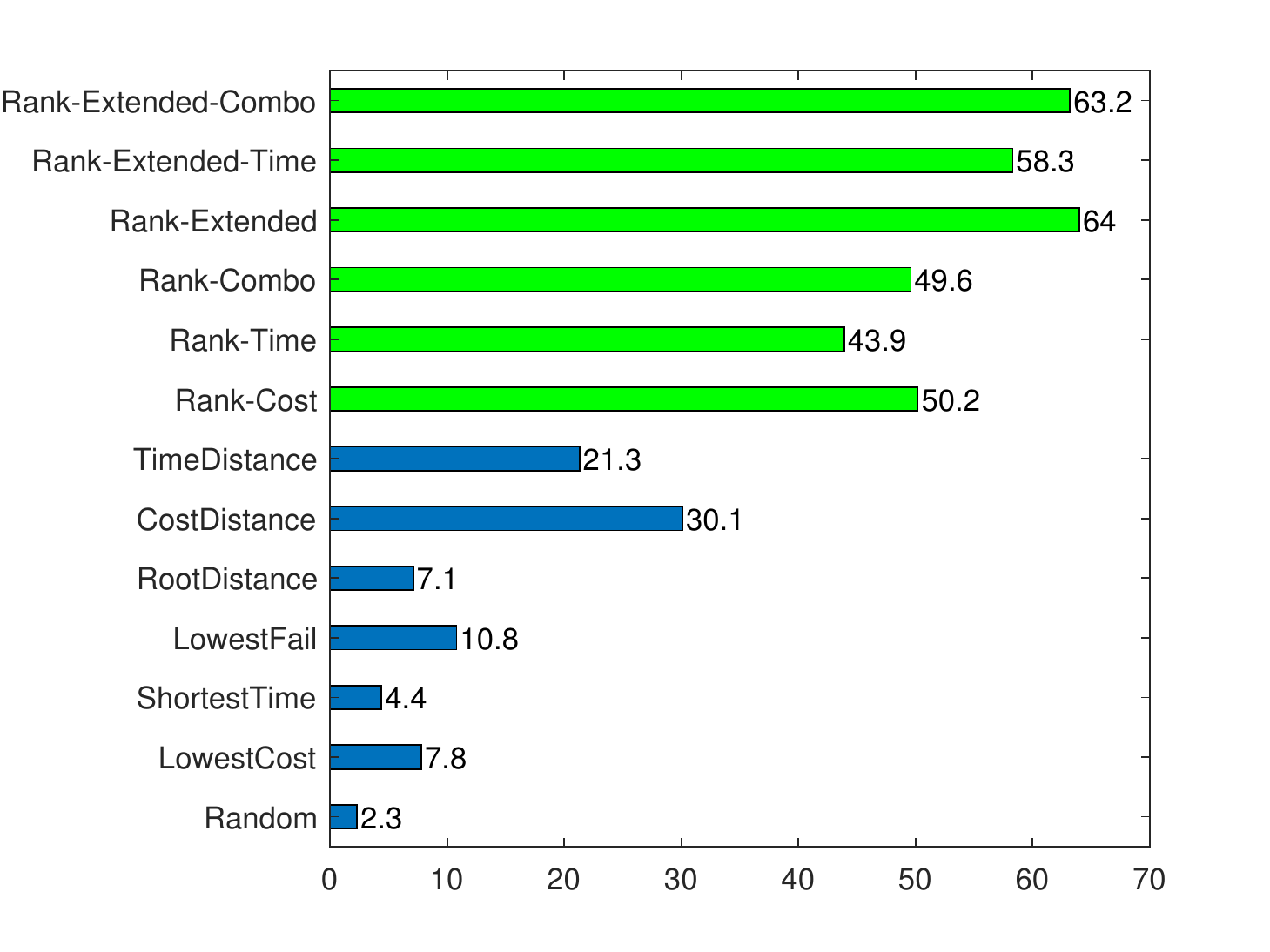}
	\includegraphics[width=0.495\textwidth]{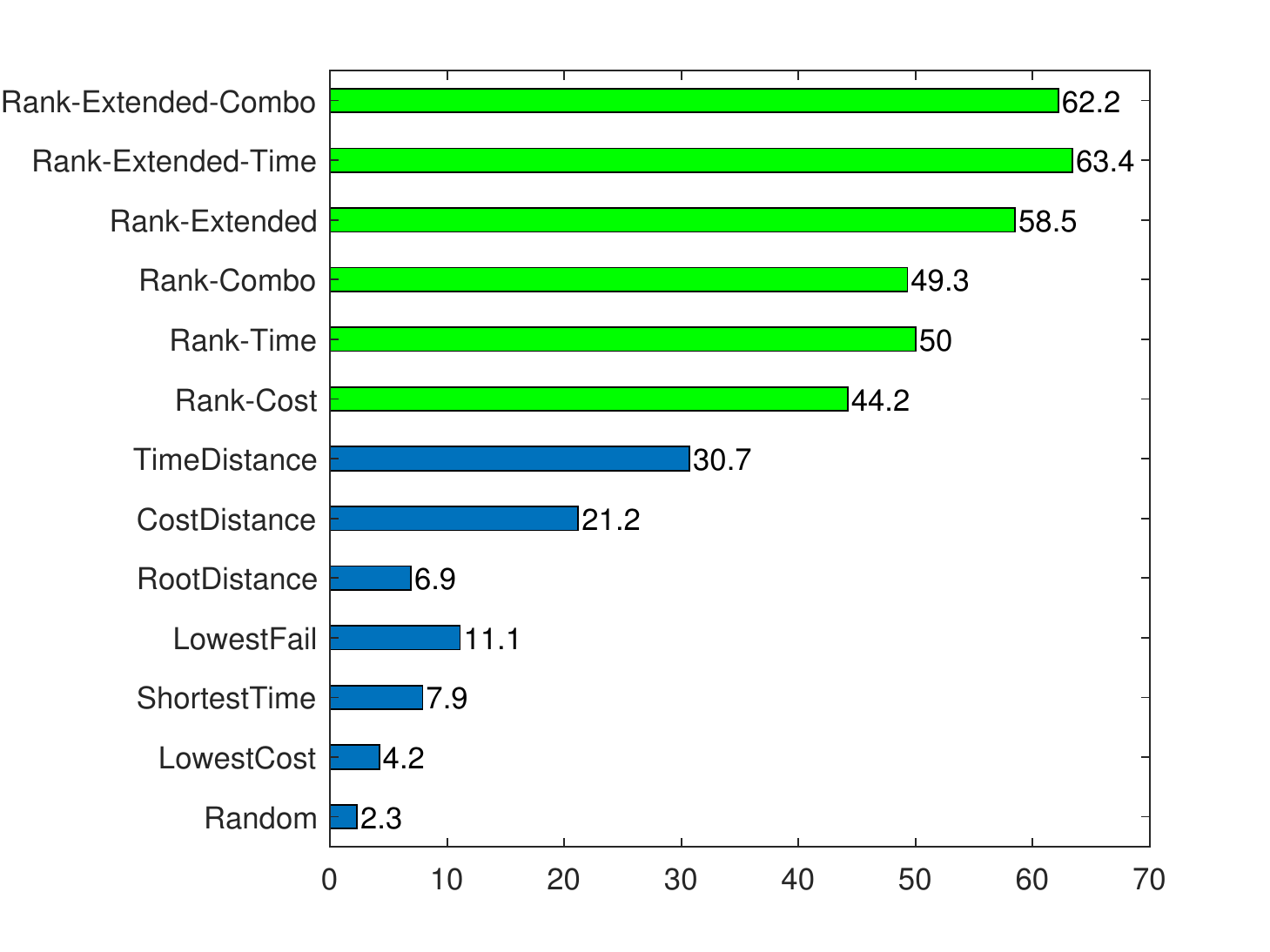}
	\caption{The percentage of cases where each heuristic achieved the best result in terms of cost (left) and time (right) in  the second setting of the evaluation.}
	\label{fig:set2_cases}
\end{figure}

Finally, as shown in Figure \ref{fig:set2_cases}, in this setting, there is a stronger correlation between the most frequent and the best performing solution.

\subsection{Impact of early termination and time overheads}
\label{sssec:other}

In this section, we discuss in detail the impact of the notion of meaningless operations which is leveraged by our extended rank-based solutions. Thus far, we have ignored all failed instances.
Remember that an execution instance is considered successful, if the root element is produced. On the contrary, when the production of the root element is not possible, i.e. due to the failed execution of some operations, the execution instance is considered unsuccessful. In such cases, its execution may be terminated earlier, as already discussed. 

In failed instances, the performance of every heuristic that does not employ early termination converges to the same cost and time duration. The extended rank-based variants, which are equipped with the early termination capability, are inherently more suited for such failed instances.
 In particular, our proposals exhibit a 13.8\% and 13.4\% improvement in terms of cost and time, respectively, in the first setting. In the second setting, the improvements are even better; 23.5\% lower cost and 23.4\% shorter time. In PDM3, there are no failed instances that lead to a benefit (we discuss this PDM further below).

Finally, the overhead time to run the heuristics is extremely low: on a Ryzen 5 3600x CPU with 16GB RAM, our rank-based heuristics take less than 1.34 milliseconds for each case of the social insurance PDM; the other heuristics are even faster and the times are negligible for the small PDM.

\subsection{Additional Discussion}
\label{sssec:additional}

In this section,  we  discuss the performance of our revised and extended rank-based solutions in comparison to the rank-based solutions, originally presented in our earlier work \cite{VarvoutasG20}. In the mortgage and social insurance PDMs, the extended techniques demonstrated significant improvements over the original variants, performing on average, 13.52\% better. In the monitoring process PDM however, the new proposals did not manage to lead to any improvements displaying exactly the same level of performance as the original flavors. This can be attributed to the particular characteristics of the structure of this PDM: (i) none of its elements are deemed as interdependent and therefore, they cannot be grouped together; (ii) this PDM also does not contain any operation pairings which comply with the notion of meaningless operations. As a result, the PDM fails to benefit from the two novel notions that were introduced in the extended rank-based techniques.

Finally, it should be noted that the performance of our approach was significantly better in the second setting. This can be attributed to the increased range of values in that setting. In particular, the higher range of values in terms of probability of failure leads to an increase in path variability, favoring our approach.

\subsection{Deviation from Optimal Solutions}
\label{sssec:opt}

\begin{figure}[tb!]
	\centering
	\includegraphics[width=0.495\textwidth]{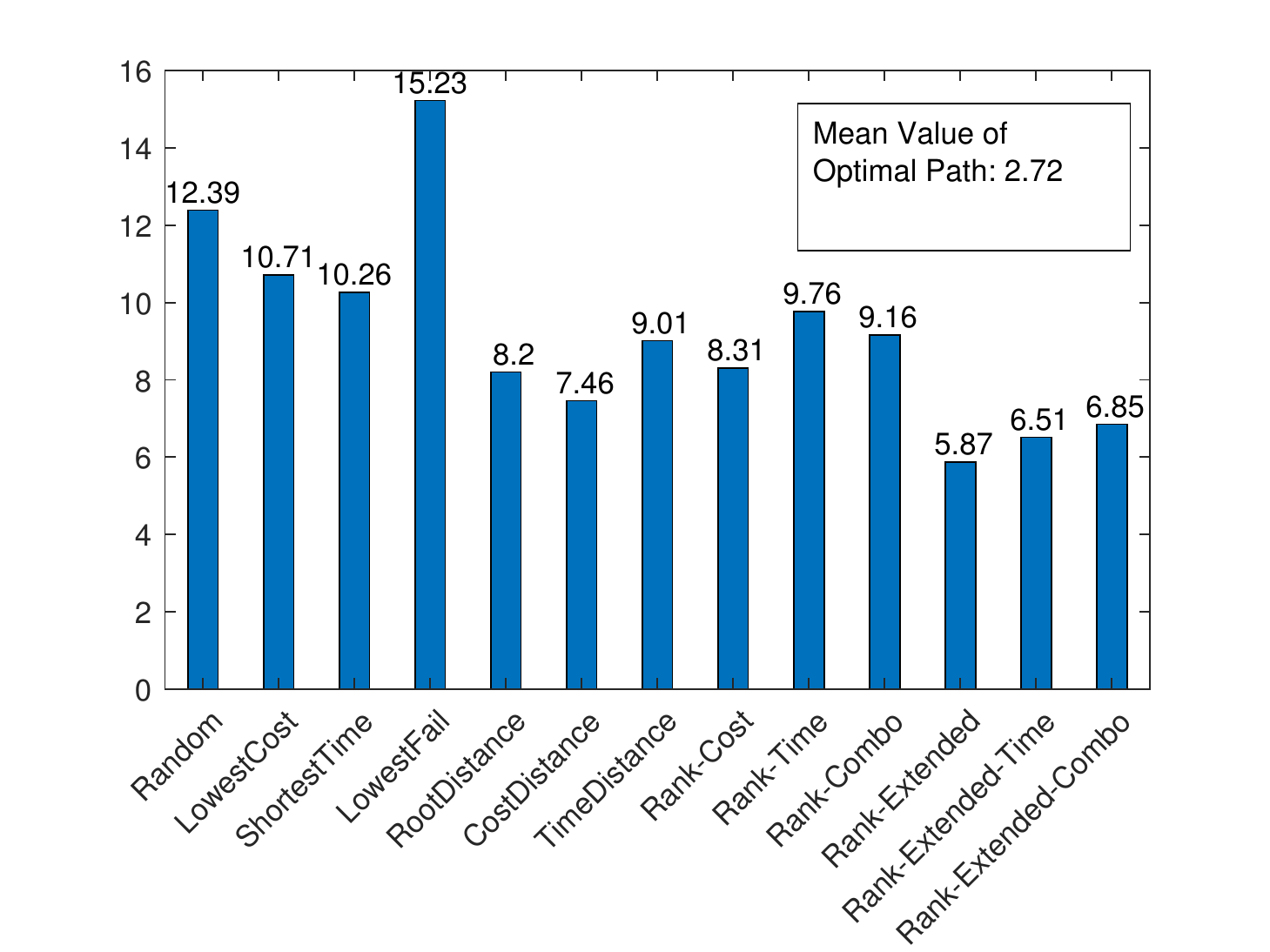}
	\includegraphics[width=0.495\textwidth]{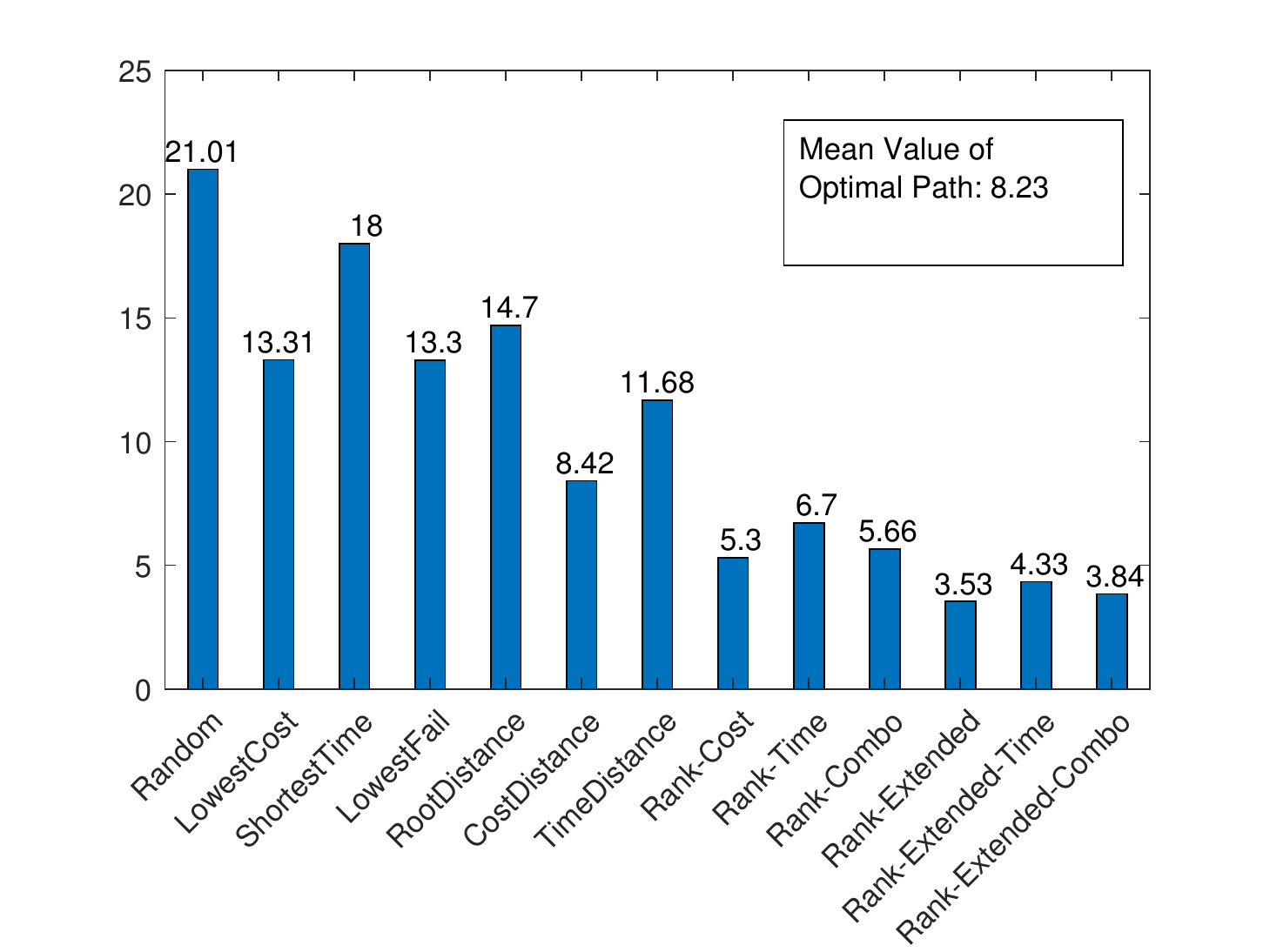}
	\caption{The average deviation of each heuristic compared to the optimal path for the first setting  (left) and the second setting (right) regarding the mortgage PDM.} 
	\label{fig:pdm1_optimal_cost}
\end{figure}

In this experiment, we rely on the notion of \emph{optimal path}, presented in Section \ref{sec:back}, to further evaluate our approach. For each execution instance, we determine the optimal path, in terms of cost and path feasibility. Then, the optimal path is directly compared to the paths discovered by each of the heuristics, for that specific instance. This comparison is conducted for the mortgage PDM only, presented in Figure \ref{fig:pdm}, which is small enough so that optimal paths can be derived.

Figure \ref{fig:pdm1_optimal_cost} presents the average  deviation from the optimal path of each heuristic, in terms of cost during the execution of the mortgage PDM. 
In the first setting, the mean cost of the optimal path is 2.74. We can see that even in this simple PDM, even the best performing rank-based heuristic deviate by  more than two times. 
However, narrower deviations are  observed in the second setting, where the best performing rank-based variant incurs a 42.9\% higher cost  on average.

The main conclusion is that although our proposed heuristics significantly outperform existing ones, there is still plenty of room for improvements. Devising fast heuristics that yield results that are closer to the optimal ones remains a challenging open issue.

\section{Related Work}
\label{sec:rw}

As stated in the introduction, an increasing amount of data-centric approaches have been developed as part of a general trend in the area of Business Process Management (BPM). Despite this recent interest, Business Process Improvement or Redesign, one of the key areas of BPM  remains relatively undeveloped in terms of automated algorithmic solutions. In a recent survey that aims to evaluate several data-centric process approaches, this lack of focus on process optimization or redesign is highlighted \cite{evaluating}. Out of the 14 data-centric methods examined, only 2 of them identify the objective of business process optimization as a motive for their development. These two methods are Product Based Workflow Design (PBWD) \cite{ReijersLA03} and its extension, Product Based Workflow Support (PBWS) \cite{pbws}, upon which we build our work.

In \cite{ComuzziV11,ComuzziVW13}, Comuzzi et. al. present an ant-colony based approach, which aims to discover the optimal path in a PDM. A key element of this approach is the distinction between the phases of path discovery and execution. However, whenever an operation executes unsuccessfully, then the entire instance is terminated. This is a limitation compared to the direct PDM execution approach of \cite{pbws} upon which we build our work, which allows to continue the execution via considering the remaining operations.  In a similar context, the work in \cite{ZXH2012} follows an ant-colony based approach to optimize the execution of a business process, represented as a Workflow Net rather than a PDM, as we do. 

Regarding task ordering in declarative process models, the work of Barba et. al. presents a recommendation approach that aims to optimally order the remaining executable activities of an ongoing process instance \cite{BarbaWVR13,BarbaWV11}. 
In a subsequent work \cite{Jimenez-RamirezW0V15}, the same authors present an approach that computes a set of potentially optimal plans for a declarative model, discarding the sub-optimal ones. This set of plans is then combined into a single, configurable, BPMN model. In all these works, the task ordering heuristics are a subset of those examined in \cite{pbws}, since the focus is mainly on producing valid execution plans rather than optimal ones according to a cost function. However, contrary to our solution, the focus is also on optimal resource scheduling.

An extensive part of recent business process literature focuses on  process model variability  \cite{Schunselaar}. For example, the work in \cite{Schunselaar} is motivated by the fact that different municipalities perform the same objective using different, but equivalent, processes. To accommodate such variability, it introduces the configurable process tree, a novel formalism capable of representing a family of process variants in a single model. Additionally, this methodology allows a specific set of process models to be selected according to a set of Key Performance Indicators (KPIs). This bears some similarity to the way PBWS and our solutions exploit the existence of alternative paths in order to optimize each case's performance. The main difference lies in the fact that, in our case, these alternatives are different paths of a single, existing PDM model, while in \cite{Schunselaar}, there is an attempt to create a model that contains alternative paths to cover all cases.
In the same context, the work in \cite{configurable} presents a set of approaches for the extraction of a configurable process model from a collection of event logs, which represents variants of a single process. Furthermore, the issue of assessing the quality of different process model configurations \cite{navigate} has also been explored. Lastly, the work in \cite{MertensGP14} presents an approach that produces a set of ranked imperative models based on an input declarative model; the imperative models are, in turn, used to generate recommendations. Such an approach focuses on flexibility and essentially checks all possible plans, thus it cannot scale.

Additionally, there are proposals considering various optimization objectives, such as the techniques in \cite{Aalst01}, where a set of heuristics for changing the relative ordering of activities is introduced. 
However, these heuristics cannot be applied to PDMs because they assume that all operations need to be executed, whereas, in PDMs, it suffices to execute a subset of operations for a process to complete.
Finally, our work relates to declarative process models
\cite{4384001,Chawla2011UserguidedDO}; e.g., our workflow design solution can be seen as a promising means to
derive executable model structures out of such declarative models although providing a complete methodology to achieve this remains an open issue.

Our rank-based solutions are inspired by solutions to optimize data analytics workflows \cite{KougkaGS18}, but they constitute novel optimization variants tailored to PDM-modeled business processed.
The work in \cite{SMA+18} highlights the benefits of integrating BPM and data analytics, by proposing the utilization of Big Data processing for the purpose of analyzing business process data. Similarly, there have been efforts to transfer the results of data analytics workflow optimization to business process workflows \cite{transfer}. 
Despite these efforts, the topic of fully exploiting the know-how in data-intensive workflows for optimizing the execution of business processes still remains in its infancy.

In the more general research area of BPM, there has been a rise in research related to the field of decision management. Many works are focusing on PDMs, the recently introduced \emph{Decision Model and Notation (DMN)} and other decision modelling approaches. The work in \cite{AaLBWR15} proposes an automated methodology that extracts the decision logic from an input PDM workflow and produces a BPMN/DMN model as an output. The work in \cite{Smedt2019} proposes an automated approach that derives decision models from enriched event logs. Moreover, the \emph{Decision Data Model (DDM)},  an extension of the PDM, was introduced in  \cite{PetruselVD11}; this work focuses on the data perspective of the decision making of a process. All these efforts are orthogonal to our work.

\section{Conclusion and Future Work}
\label{sec:concl}

This work focuses on business processes modelled according to the declarative PDM paradigm. PDM is a powerful modeling approach for business processes that aim to produce informational products, e.g., whether to grant an approval to a specific admission request, without specifying the workflow execution plan.  The contribution of this work is that it proposes and evaluates novel heuristics for yielding optimized workflow designs, i.e., execution plans, in terms of execution cost and time  on a case-by-case basis.  Our solution is inspired by and extends optimization solutions in query processing and workflows for data analytics. Its main novelty lies in the fact that, when choosing the next operation to be executed, it takes into consideration both the 
 probability to produce the root element and the cost of the operation. We also consider explicitly operation inter-dependencies and discuss implementation issues. The evaluation is based on three real-world PDM examples and tens of thousands of random metadata instantiations for these PDMs following realistic probability distributions. Our experimental results show that, although in specific cases different heuristics may perform better, our proposals consistently outperform existing solutions. They yield, on average, 16.5\% lower cost and 22.6\% shorter time, when the metadata are based on real-world data. For random metadata following a uniform distribution within a predefined range, the improvements are even higher.


This work aims to open a new avenue in which the execution of  declarative process models is optimized. However, to achieve this goal, two main directions need to be explored in depth in the future. Firstly, our approach needs to cover additional declarative modelling approaches, such as \cite{4384001} and also consider processes with multiple outputs. Secondly, and more importantly, a complete cost and time optimization solution should also consider aspects, such as  resource utilization, parallel task execution and smaller deviations from optimal solutions. Especially with regards to the issue of resource allocation, the current solution presented in this work implicitly assumes that all resources required for an operation execution are always available; extending our algorithms to account for resource unavailability, non-deterministic behavior, waiting times and fatigue is essential to increase the applicability of the proposal in real-world settings.
Additionally, we aim to expand the scope of our work by investigating the composition of activities, similarly to \cite{VANDERAA2016162},  which impacts on the PDMs themselves apart from their execution only.

\emph{Acknowledgment.} The research work was supported by the Hellenic Foundation for
Research and Innovation (H.F.R.I.) under the ``First Call for H.F.R.I.
Research Projects to support Faculty members and Researchers and
the procurement of high-cost research equipment grant” (Project
Number:1052, Project Name: DataflowOpt).

\appendix
\section{PDM original metadata}

In this section, we present in Table \ref{tab:table-data-ap1} the original data of the product data models that were used in the first setting of the evaluation in Section \ref{sssec:fs} of our work.

\begin{table}[tbp!]
	\begin{center}
	\resizebox*{0.85\textwidth}{\dimexpr\textheight-\lineskip\relax}{
		\begin{tabular}{|c | c | c | c | c|} 
			\hline
			\bf{ID} & \bf{Output} & \bf{Input} & \bf{Cost} & \bf{Failure Probability}\\ [0.5ex] 
			\hline
			Op01 & i01 & i25, i27 & 0.0 & 0.0 \\
			\hline
			Op02 & i02 & i25, i37 & 0.0 & 0.0\\
			\hline
			Op03 & i03 & i33, i37 & 0.0 & 0.0\\ 
			\hline
			Op04 & i04 & i33, i37 & 0.0 & 0.0\\ 
			\hline
			Op05 & i05 & i37, i45 & 0.0 & 0.0\\ 
			\hline
			Op06 & i06 & i21, i37 & 0.0 & 0.0\\ 
			\hline
			Op07 & i07 & i24, i37 & 0.0 & 0.0\\ 
			\hline
			Op08 & i08 & i23, i37 & 0.0 & 0.0\\ 
			\hline
			Op09 & i24 & i39, i37 & 0.0 & 0.0\\ 
			\hline
			Op10 & i10 & i13, i14, i34, i37, i42 & 0.0 & 0.0\\ 
			\hline
			Op11 & i11 & i31 & 0.6 & 0.0\\ 
			\hline
			Op12 & i15 & i16 & 0.0 & 0.003\\ 
			\hline
			Op13 & i15 & i17 & 0.0 & 0.997\\ 
			\hline
			Op14 & i15 & i16, i17 & 0.0 & 0.0\\ 
			\hline
			Op15 & i16 & i25, i30, i35, i36, i44 & 5.61 & 0.0\\ 
			\hline
			Op16 & i17 & i25, i30 & 6.1 & 0.0\\ 
			\hline
			Op17 & i18 & i01 & 0.0 & 0.991\\ 
			\hline
			Op18 & i18 & i02 & 0.0 & 0.987\\ 
			\hline
			Op19 & i18 & i08 & 0.0 & 0.984\\ 
			\hline
			Op20 & i18 & i09 & 0.0 & 0.998\\ 
			\hline
			Op21 & i18 & i10 & 0.0 & 0.932\\ 
			\hline
			Op22 & i18 & i11  & 0.0 & 0.981\\ 
			\hline
			Op23 & i18 & i15 & 0.0 & 0.79\\ 
			\hline
			Op24 & i18 & i09, i11, i15 & 0.0 & 0.0\\ 
			\hline
			Op25 & i28 & i25, i37 & 0.0 & 0.0\\ 
			\hline
			Op26 & i29 & i25, i30, i35, i36 & 0.0 & 0.0\\ 
			\hline
			Op27 & i30 & i32, i37, i43 & 0.0 & 0.0\\ 
			\hline
			Op28 & i31 & i29, i40, i48 & 0.0 & 0.0\\ 
			\hline
			Op29 & i32 & i01, i02, i03, i04, i05, i06, 
			i07, i08, i10, i27, i28 & 0.0 & 0.0\\ 
			\hline
			Op30 & i34 & i36, i37, i41 & 4.2 & 0.0\\ 
			\hline
			Op31 & i40 & i39, i41 & 0.3 & 0.0\\ 
			\hline
			Op32 & i42 & i47 & 0.3 & 0.0\\ 
			\hline
			Op33 & i43 & i39, i49 & 0.6 & 0.0\\ 
			\hline
			Op34 & i13 & - & 0.08 & 0.0\\ 
			\hline
			Op35 & i14 & - & 0.08 & 0.0\\ 
			\hline
			Op36 & i21 & - & 0.0 & 0.0\\ 
			\hline
			Op37 & i23 & - & 0.67 & 0.0\\ 
			\hline
			Op38 & i24 & - & 0.0 & 0.0\\ 
			\hline
			Op39 & i25 & - & 0.0 & 0.0\\ 
			\hline
			Op40 & i27 & - & 0.08 & 0.0\\ 
			\hline
			Op41 & i33 & - & 0.0 & 0.0\\ 
			\hline
			Op42 & i35 & - & 0.0 & 0.0\\ 
			\hline
			Op43 & i36 & - & 1.0 & 0.0\\ 
			\hline
			Op44 & i37 & - & 1.67 & 0.0\\ 
			\hline
			Op45 & i39 & - & 0.17 & 0.0\\ 
			\hline
			Op46 & i41 & - & 0.0 & 0.0\\ 
			\hline
			Op47 & i44 & - & 0.0 & 0.0\\ 
			\hline
			Op48 & i45 & - & 0.0 & 0.0\\ 
			\hline
			Op49 & i47 & - & 0.33 & 0.0\\ 
			\hline
			Op50 & i48 & - & 0.0 & 0.0\\ 
			\hline
			Op51 & i49 & - & 0.0 & 0.0\\ 
			\hline
		\end{tabular}
		}
		\caption{\label{tab:table-data-ap1}Operation attributes of the social insurance PDM.}
	\end{center}
\end{table}

\clearpage
Next, in Table \ref{tab:table-data-ap2}, we present the data of the monitoring process product data model.

\begin{table}[tbp!]
	\begin{center}
	\footnotesize
		\begin{tabular}{|c | c | c | c | c | c|} 
			\hline
			\bf{ID} & \bf{Output} & \bf{Input} & \bf{Cost} & \bf{Quality} & \bf{Failure Probability}\\ [0.5ex] 
			\hline
			Op01 & i1 & i2, i3 & 0.0 & 1.0 & 0.0 \\
			\hline
			Op02 & i3 & i4 & 0.6 & 0.3 & 0.0\\
			\hline
			Op03 & i3 & i4, i5 & 0.8 & 0.8 & 0.0\\ 
			\hline
			Op04 & i3 & i5 & 0.1 & 0.5 & 0.0\\ 
			\hline
			Op05 & i2 & i6 & 0.1 & 0.2 & 0.0\\ 
			\hline
			Op06 & i2 & i7 & 0.8 & 1 & 0.0\\ 
			\hline
			Op07 & i2 & i8, i10 & 0.8 & 1.0 & 0.0\\ 
			\hline
			Op08 & i2 & i8, i11 & 0.6 & 0.7 & 0.0\\ 
			\hline
			Op09 & i2 & i9, i10 & 0.6 & 0.7 & 0.0\\ 
			\hline
			Op10 & i2 & i9, i11 & 0.5 & 0.5 & 0.0\\ 
			\hline
			Op11 & i4 & - & 0.2 & 1.0 & 0.1\\ 
			\hline
			Op12 & i5 & - & 0.7 & 1.0 & 0.3\\ 
			\hline
			Op13 & i6 & - & 1.0 & 1.0 & 0.3\\ 
			\hline
			Op14 & i7 & - & 0.8 & 1 & 0.1\\ 
			\hline
			Op15 & i8 & - & 0.5 & 1 & 0.37\\ 
			\hline
			Op16 & i9 & - & 0.3 & 1 & 0.3\\ 
			\hline
			Op17 & i10 & - & 0.3 & 1 & 0.3\\ 
			\hline
			Op18 & i11 & - & 0.1 & 1 & 0.2\\ 
			\hline
	    \end{tabular}
		\caption{\label{tab:table-data-ap2}Operation attributes of the monitoring process PDM.}
	\end{center}
\end{table}

\end{document}